# Exceeding octave tunable Terahertz waves with zepto–second level timing noise

Rubab Amin[1], James Greenberg[1], Brendan Heffernan[1], Tadao Nagatsuma[2], and Antoine Rolland[1]

[1]IMRA America, Inc., Boulder Research Lab, 1551 S. Sunset St. Suite C, Longmont, Colorado 80501, USA

[2]Graduate School of Engineering Science, Osaka University, Osaka 560-8531, Japan

## Abstract

Spectral purity of any millimeter wave (mmW) source is of the utmost interest in low-noise applications. Optical synthesis via photomixing is an attractive source for such mmWs, which usually involves expensive spectrally pure lasers with narrow linewidths approaching monochromaticity due to their inherent fabrication costs or specifications. Here, we report an alternative option for enhancing the spectral purity of inexpensive semiconductor diode lasers via a self–injection locking technique through corresponding Stokes waves from a fiber Brillouin cavity exhibiting greatly improved phase noise levels and large wavelength tunability of ~1.8 nm. We implement a system with two self–injected diode lasers on a common Brillouin cavity aimed at difference frequency generation in the mmW and THz region. We generate tunable sub-mmW (0.3 and 0.5 THz) waves by beating the self – injected two wavelength Stokes light on a uni-travelling carrier photodiode and characterize the noise performance. The sub-mmW features miniscule timing noise levels in the zepto-second (zs.Hz$^{-½}$) scale outperforming the state of the art dissipative Kerr soliton based micro-resonator setups while offering broader frequency tunability. These results suggest a viable inexpensive alternative for mmW sources aimed at low-noise applications featuring lab-scale footprints and rack-mounted portability while paving the way for chip-scale photonic integration.

## Introduction

Spectrally pure and tunable millimeter wave (mmW) signals (e.g. 100 – 600 GHz) will enable progress in a growing number of applications, such as deployment of the digital cellular network 5G [1] and prospective 6G [2], vehicular communications [3], and scientific research amongst which radio-astronomy [4, 5], ultra-precise atomic and molecular spectroscopy [6, 7], particle accelerators [8], and remote sensing [9], are just a few examples. In the past century, microwave technologies have allowed the generation of such wave relying on the frequency multiplication in several mmW and terahertz bands using a microwave reference (typically operating up to 10 GHz). However, this technique suffers from its inherent nature of not only multiplying the frequency but also its phase noise, ultimately degrading the spectral purity of the generated mmW.

In the last few decades, the performance of mmW sources has greatly accelerated with photonics technology. Photomixing two optical wavelengths in the telecom C band or an optical pulse train with high repetition rate frequency beating on a uni-travelling carrier photodetector (UTC-PD) emitting in free space or in a waveguide is a promising approach to outstrip the electronic counterparts. In fact, the use of an optical pulse train at 300 GHz as an optical frequency divider of an optically carried multi-THz reference has led to noise metrology in the mmW domain obtaining unprecedented levels [10]. However, the rigid tunability of an optical pulse train makes it difficult to cover wide bands in the mmW domain. Therefore, beating two widely tunable optical wavelengths on a fast photodiode is a powerful technique. The challenge of such approach lies in finding ways to strongly correlate the two optical wavelengths to leverage common mode rejection at photodetection. When two optical signals bearing disparate wavelengths oscillate in a common resonator, such common mode rejection becomes realistic [11]. Over the years, this has been implemented using solid-state lasers [12], on-chip semiconductor lasers [13], and fiber lasers [14].

Continuous wave lasers with extreme spectral purity at fast Fourier frequencies have been made possible by generating an optical Stokes wave from a pump laser through stimulated Brillouin scattering (SBS) in fiber ring resonators, and even in waveguide-based microresonators [15-18]. One technical challenge in such lasers is to suppress mode-hopping. When SBS is generated through resonant pumping of a microresonator, mode-hopping suppression is guaranteed as the free spectral range is much higher than the Brillouin gain bandwidth whereas the quality factor is curtailed and the tunability is also strongly limited. In contrast, a long fiber cavity stimulated with a non-resonant pump laser, leads to the generation of a Stokes wave widely tunable by simply tuning the pump laser wavelength. However, this comes with frequent mode-hopping as the free spectral range becomes much smaller than the Brillouin gain bandwidth. Mode-hopping can be cancelled in long fiber cavities with an optoelectronic phase locked loop [19], and, as recently demonstrated, using the optical injection of the Stokes wave in the pump laser generating an optical wave with a spectral purity that remains unmatched at Fourier frequencies above 10 kHz [20]. As a consequence, it becomes interesting to generate two Stokes waves simultaneously to achieve a spectrally pure and tunable mmW signal on a fast photodiode.

In this report, we have developed a novel millimeter wave source tunable from a few tens of GHz up to 0.5 THz based on two highly correlated optical lines generated with a common 75 m long fiber ring cavity through an SBS process with two commercially available, non-resonant semiconductor laser diodes to generate ultralow noise millimeter to terahertz waves. Mode hopping suppression is realized with self-injection of the Stokes wave back into each corresponding pump laser. Not only does this allow for pure single mode operation for each laser but it also allows the copying of the cavity intrinsic phase noise into the phase noise of corresponding pump lasers. With merely off-the-shelf components, we demonstrate that such a scheme allows for linewidth reduction of a pump laser by approximately four orders of magnitude (~$10^4$) while maintaining sub-100 Hz linewidth over the tuning range of a "noisy" diode laser. We implement a system containing two self – injected diode lasers to produce beat notes in the millimeter wave to THz region of the spectrum spanning across 180 – 520 GHz with greatly reduced noise levels with noise floors reaching zepto-second scales (approximately 70 zs.Hz$^{-\frac{1}{2}}$). Phase noise analysis, realized with original photonic setups relying on a fiber delay line, reveals that the demonstrated tunable millimeter-wave oscillator reaches an unprecedented timing noise level in the zepto-second regime with sub-100 Hz linewidth promising to unlock limitations in applications that requires ultimate noise performance in the millimeter- and sub-millimeter-wave domains. Our results suggest a viable, inexpensive alternative for mmW sources aimed at low-noise applications featuring lab-scale footprints occupying only a couple of optical breadboards confined within a small portion of an optical table, and a useful rack-mountable portable alternative to the costly and sizeable options usually utilized for low-noise mmW synthesizers. Additionally, size and power consumption reduction efforts, coupled with state of the art lithography and associated nanofabrication technologies perfected over the recent years, have led to the emergence of foundry based chip-scale integrated photonics. As chip-scale solutions for narrow linewidth and low-noise laser sources are becoming an apparent requirement for next generation integrated photonic applications, it is worthwhile to mention here that the approach described herein for linewidth reduction and performance enhancement of a semiconductor diode laser thereto achieving low-noise THz waves extends beyond just fiber based setups and paves the way for such chip-scale integration utilizing the self – injection scheme outlined here.

## Difference Frequency Generation Scheme based on Self – Injected Laser Diodes

Our source consists of two continuous wave (CW) semiconductor distributed feedback diode lasers (Eblana Photonics, EP1550 Discrete-Mode Series) operating in the telecommunication C-band with operating wavelengths, $\lambda_1$ and $\lambda_2$ centered around 1550 nm and 1552 nm, respectively; for mmW to THz wave generation through utilizing beat notes based on their operating wavelength differences when they beat on a fast enough photodiode together (Fig. 1a, b). The employed lasers incorporate no optical isolators to facilitate optical injection and feature relatively broad linewidths specified roughly around 800 kHz. Both laser outputs are combined with a fiber coupler and the combined two wavelength $(2 - \lambda)$ light subsequently non-resonantly pumps a 75 m long fiber Brillouin ring resonator. The Brillouin cavity features a 95:5 fiber coupler and is set for non-resonant pumping through the use of a fiber circulator, i.e. only the backwards propagating Stokes wave is resonant on the cavity. The Brillouin cavity is placed inside a vacuum chamber with temperature and vibration control to stabilize the performance independently, free from any external environmental factors. The corresponding acoustic damping from the silica fiber redshifts the $2-\lambda$ light output by the amount of the resultant Brillouin shifts in fiber, $\nu_{B1}$ and $\nu_{B2}$ at the corresponding operating wavelengths, $\lambda_1$ and $\lambda_2$, respectively. The Brillouin frequency shift, $\nu_B$ depends on the material composition and to some extent the temperature and pressure of the fiber [21]. Note, all experiments are carried out in room temperature. The Brillouin fiber ring Stokes radiation linewidth can be several orders of magnitude lower than that of the incident pump beam [22]. The Brillouin cavity operates as a nonlinear narrow bandwidth filter which passes radiation at the Stokes wavelength only [23]. Contrary to lasers with population inversion, the spontaneous Brillouin scattering fundamentally limits the degree of monochromaticity of the Stokes radiation as opposed to spontaneous emission in recombination-generation schemes [15]. As such, both CW lasers generate disparate SBS lasers on the same cavity based on corresponding Stokes waves which are exactly separated by the pump laser operating wavelengths. The Stokes output is next passed through an electro-optic modulator (EOM) in order to transfer the Brillouin shifted Stokes light back into the frequency of the corresponding pumps depending upon the operational frequency of the EOM, which is set to match the relevant Brillouin shifts, $\nu_{B1}$ and $\nu_{B2}$. A Mach–Zehnder intensity modulator can be utilized as the necessary EOM in this setup the output of which is subsequently feedback to the pump lasers through the aforementioned fiber coupler. A single EOM can be driven in dual-tone modulation arrangement with corresponding frequency synthesizers to facilitate both the required Brillouin shift frequencies as both such frequencies lie in rather close vicinity of each other (~10.8 GHz), the resultant output from the EOM for a single wavelength SBS radiation contains sidebands generated from operation which can be expressed in frequencies as $\nu_{s1,2} \pm mf_{EOM1,2} \equiv \nu_{LD1,2} \pm (m-1)\nu_{B1,2}$; where $\nu_{LD1,2}$ is the CW output frequency of either pump laser, and $m$ is an integer corresponding to the order of the EOM sidebands. In our operation, we only require the first blue-shifted sideband from modulation, $m=1$ to be injected back into the pump laser; however, the remaining sidebands from EOM operation do not need to be filtered out as the corresponding pump frequency, $\nu_{LD1,2}$ is not affected by large detunings. The resultant sidebands from EOM operation are also easily tuned in power using a DC input to the EOM. As such, each diode laser can be self–injection locked and improve the corresponding output efficiency iteratively in the loop until a thermal equilibrium of the Brillouin cavity noise floor is reached. The $2-\lambda$ Stokes radiation from the Brillouin cavity is the crucial output of our system manifesting two SBS lasers ensued from a single Brillouin cavity which generate the desired mmW frequency on a fast photodiode.

Investigation into the phase noise of each of the SBS output independently, i.e. the out of loop signals, with a delayed self-heterodyne measurement setup revealed greatly reduced noise levels featuring roughly -53 dBc/Hz, -85 dBc/Hz and -110 dBc/Hz levels at Fourier frequencies of 1 kHz, 10 kHz and 100 kHz, respectively (see methods and supplementary information). The self-heterodyne measurement is carried out at an intermediate frequency of 80 MHz and a delay fiber length of 87 m single mode fiber (SMF), which is more than the coherence length of each of the pump lasers. Note, the optical coherence time, usually construed as the time for the phase to diffuse over a radian on average [24], is much greater for the SBS laser than for the pump laser [15].

Furthermore, the CW diode lasers are tunable in wavelength via drive current, $I_{LD1,2}$ alterations. This necessarily indicates feasibility of wavelength tuning the corresponding SBS lasers also. The pumps, and consequently SBS, lasers are widely tunable by approximately 1.8 nm within the lasing regime of the pumps. The pump lasers feature a threshold current for lasing at about 14 mA and are rated for safe maximum operation at 120 mA. Accordingly, we investigate the corresponding performance of the out of loop SBS lasers by tuning the pump lasers with $I_{LD1,2}$ ranging from approximately 20 mA to about 117 mA in 10 mA steps. This broad tunability causes variation in the corresponding Brillouin shift of each of the CW lasers in fiber according to the operating wavelength and precise control of the modulation frequency is required. As such, we utilize a frequency synthesizer with precise external control, decent noise characteristics, and significantly reduced footprint compared to usual sizable options (Valon 5019). We investigate the phase noise power spectral densities (PSD) of the out of loop SBS radiation at disparate wavelengths by tuning each of the pump laser current and simultaneously adjusting the power and frequency of the synthesizer driving the EOM to account for the corresponding Brillouin shift in fiber for the operational wavelength (Fig. 1c). Experimental results exhibit similar trends in the phase noise PSD for all varied wavelengths illustrating the robustness

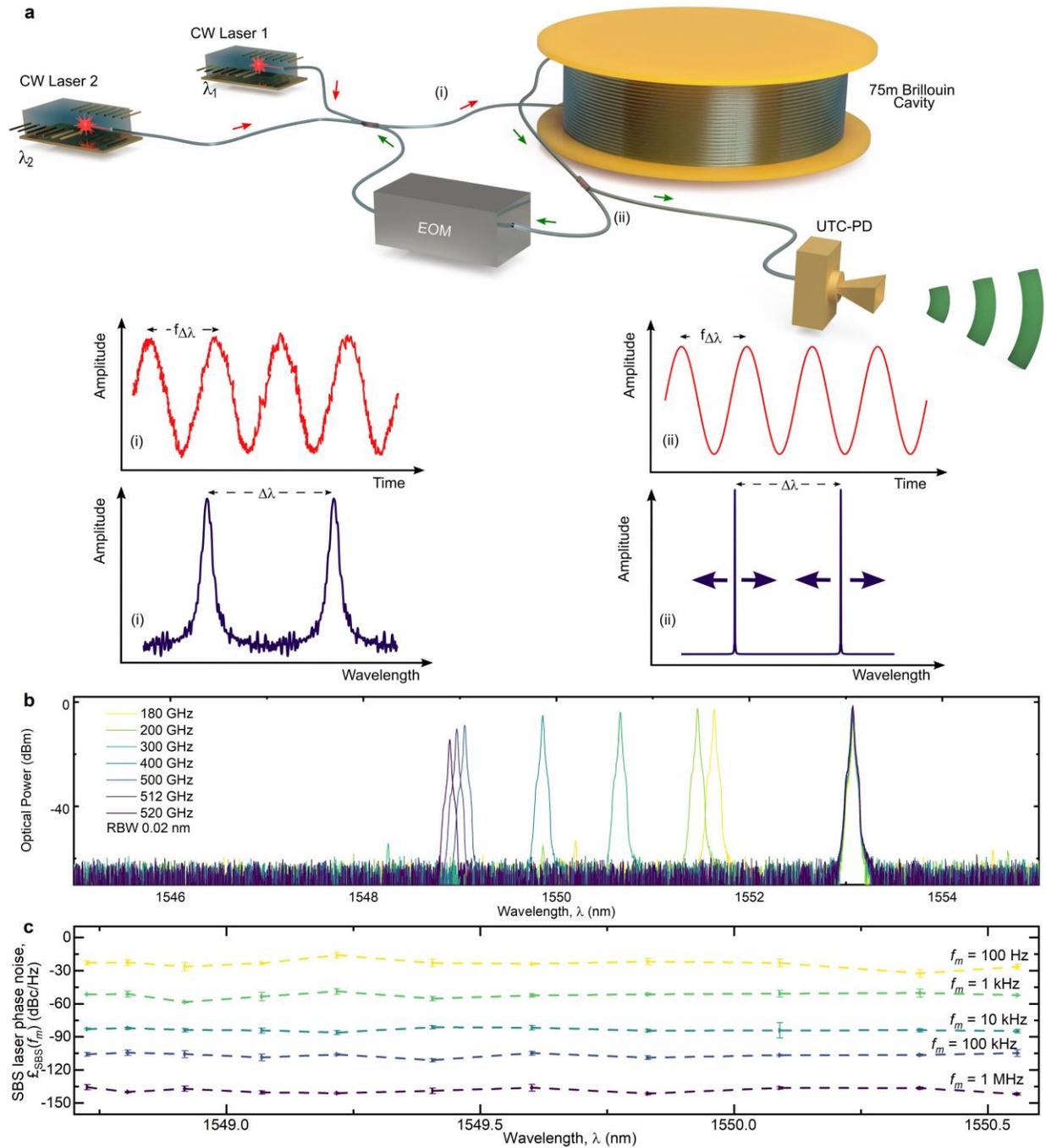

**Fig. 1: Stimulated Brillouin scattered (SBS) self – injection locking based two wavelength (2 − λ) difference frequency generation scheme. (a)** Conceptual schematic diagram of the 2 − λ setup. Two CW pump lasers are combined and sent to a non-resonantly pumped Brillouin cavity where the backwards propagating Stokes light (i.e. the two SBS lasers) is collected from the cavity and fed to an electro-optic modulator (EOM) driven dual-tone at the corresponding Brillouin shift frequencies in fiber depending on respective operating wavelengths. The EOM generated sidebands are subsequently fed back into the pump lasers to achieve the self – injection criteria. The 2 − λ Stokes light are tapped off to beat on a fast uni-travelling carrier photodiode (UTC-PD) and produce millimeter wave (mmW) to THz radiation manifesting the difference in frequency between the pump CW lasers. **(i)** Only the pump CW lasers (i.e. input to the Brillouin cavity, marked with (i)) can produce noisy mmW radiation owing to inherent relatively larger linewidth and inferior noise properties of the pump lasers, whereas **(ii)** Self – injected pump lasers based on

their corresponding Stokes wave refined SBS lasers avail spectrally pure low-noise mmW synthesis. (i) and (ii) are conceptual sketches to aid reader understanding, not actual data. **(b)** Optical spectra of different self – injected $2 - \lambda$ Stokes light in the mmW to THz regime ranging from 180 GHz – 520 GHz manifesting over an octave of frequency tunability of the demonstrated scheme. The resolution is limited by the optical spectrum analyzer (OSA) resolution bandwidth (RBW) of 0.02 nm. **(c)** SBS laser phase noise, $\mathcal{L}_{SBS}(f_m)$ in dBc/Hz corresponding to wavelength, $\lambda$ (nm) tuning of the pump laser via drive current, $I_{LD}$ (mA) variation for selected Fourier frequencies, $f_m$ of interest, i.e. 100 Hz, 1 kHz, 10 kHz, 100 kHz, and 1 MHz. Each SBS laser exhibits highly reduced and relatively unchanging levels of phase noise properties corresponding to their respective Fourier frequencies, especially in the larger Fourier frequency regions.

of the self – injection scheme for each CW laser over the wide tuning range of the corresponding diodes. Results confirm distinct Fourier frequencies featuring relatively monotonous constant phase noise levels, even with the large (~1.8 nm) tunable wavelength operation around each center wavelength, $\lambda_1$ and $\lambda_2$; namely 1550 nm and 1552 nm (Fig. 1c). It is apparent from the wavelength tuning results, as one moves further away from the carrier in Fourier frequency; not only does the phase noise level decrease significantly, but also the relative fluctuations in any chosen phase noise level corresponding to its Fourier frequency, decreases considerably; i.e. the fluctuations from the mean of the corresponding trend-line decrease. This reflects the fact that ranges in smaller Fourier frequencies (~100 Hz) are in the similar magnitude of each of the SBS linewidth itself (~40 Hz, see methods and supplementary information), which complicates phase noise measurements in that range and such results are not to be construed as optimal at face value.

We notice, even with the varying Brillouin shift in fiber corresponding to different wavelength of operation, over 0.5 nm of wavelength tuning is possible keeping the synthesizer frequency fixed with higher output power (8 dBm). The pump-Stokes beat is able to remain locked at 10.876 GHz for pump drive current ranging from 48.6 mA – 78.8 mA, which corresponds to wavelength tuning range of 1549.054 nm – 1549.598 nm; spanning 0.544 nm analogous to a remarkable ~68 GHz of tunability within the self–injection locked state (see supplementary video). Tuning wavelength beyond this range forces the Brillouin shift in fiber to change exceedingly, which cannot be mitigated by the locking condition without having to adjust for the synthesizer frequency. This revelation confirms another degree of robustness for the demonstrated self – injection scheme.

## $2 - \lambda$ Difference Frequency Noise and Stability Characterization

Next, we focus on generating a beat note between the corresponding resultant Stokes waves simultaneously generated from two self – injected lasers utilizing a single Brillouin cavity. Both CW laser outputs are passed through couple of 90:10 fiber power splitters (Fig. 2a). Both of the greater power outputs are then combined and fed into an erbium doped fiber amplifier (EDFA) via another 50:50 fiber coupler. The output of the EDFA is subsequently fed into the 75 m long fiber Brillouin cavity. The output light containing the $2 - \lambda$ Stokes waves are split into two different channels and passed through an EOM in each channel; corresponding to match the respective Brillouin shifts, $\nu_{B1}$ and $\nu_{B2}$ independently in order to transfer the Brillouin shifted stokes light into the frequency of the respective pumps (Fig. 2a). We drive the EOMs with two independent frequency synthesizers (Valon 5019) followed by RF amplifiers in the Brillouin shift frequency range (i.e. near 10.8 GHz). Mach-Zehnder intensity modulators are used as EOMs in our setup. The output from each of the EOMs is subsequently attenuated with a variable optical attenuator (VOA) for precise control of the optical power in each channel, which are then injected back into their respective CW lasers via the lower power port of the initial fiber coupler. The output from the Brillouin cavity containing the $2 - \lambda$ Stokes light is the output of this system manifesting two SBS lasers, which in turn, also refine and improve the noise and stability properties of the corresponding beat frequency, i.e. the $2 - \lambda$ difference ($\Delta\lambda$) frequency.

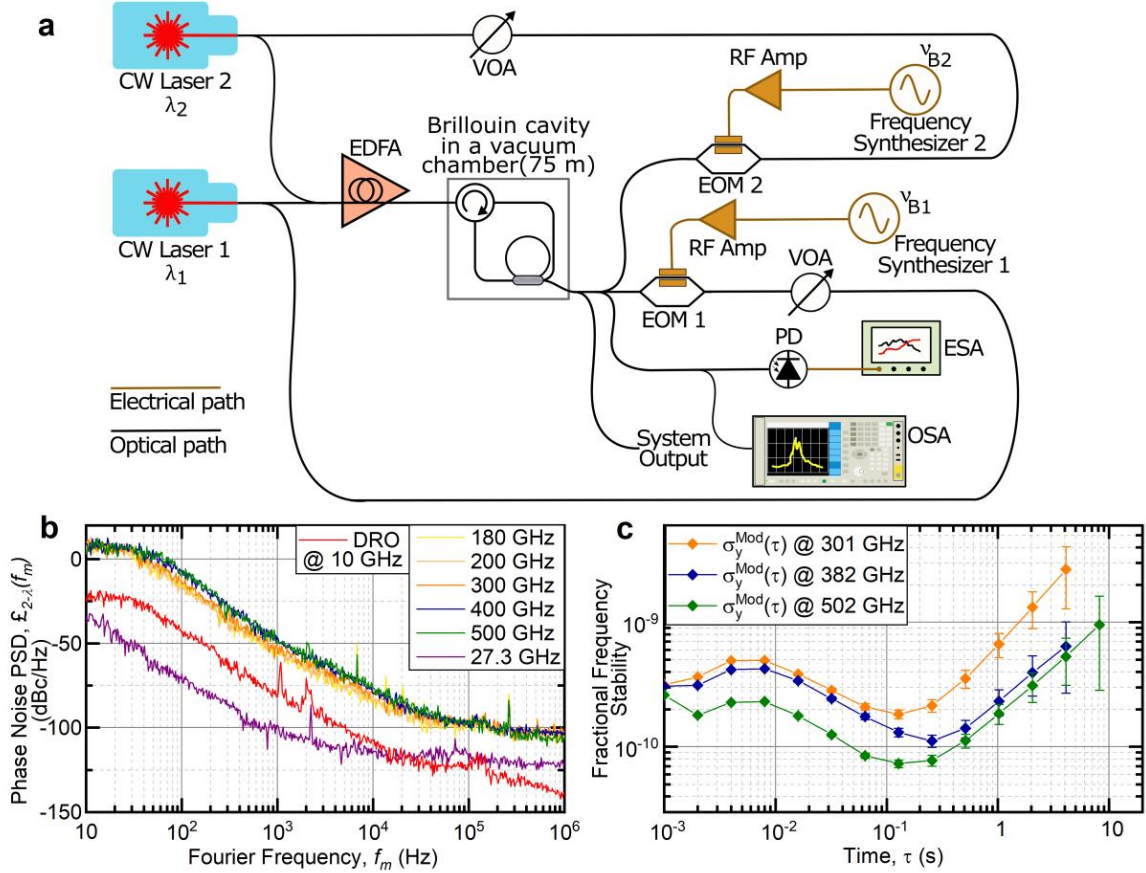

**Fig. 2: Self – injection locking based two wavelength (2 − λ) difference frequency generation scheme along with noise and stability characteristics for the different 2 − λ signals.** (a) Two CW pump lasers are amplified with an Erbium doped fiber amplifier (EDFA) and sent to a non-resonantly pumped Brillouin cavity. The backwards propagating Stokes light is collected from the cavity and split into 2 channels containing an electro-optic modulator (EOM) driven at the Brillouin shift frequency in fiber depending on respective operating wavelength with a frequency synthesizer and RF amplifier on each channel. The EOMs generate sidebands according to the synthesizer frequency, which are subsequently fed back into the pump lasers separately via another set of variable optical attenuators (VOA) for controlling the feedback power. The 2 different Stokes light are split to heterodyne on a fast photodiode (PD) and observed with an electrical signal analyzer (ESA) and optical signal analyzer (OSA) simultaneously. The light containing the two Stokes waves from the Brillouin cavity is the output of the system generating the difference in frequency between the corresponding pumps in a heterodyne beat. (b) Phase noise power spectral density of the 2 − λ Stokes beat-note directly measured at 27.3 GHz and higher bandwidth beat notes ranging from 180 GHz – 500 GHz measured with an EO comb based setup driven by a dielectric resonant oscillator (DRO) at ~10 GHz. The DRO noise is additive to the beat frequency base noise due to the employed measurement system at high beat frequencies. (c) Fractional frequency stability of different 2 − λ difference beat frequencies, which get progressively more stable at larger differences of the pump wavelengths.

In order to characterize the noise characteristics of the generated 2 − λ frequencies, we first beat both Stokes waves on a fast photodiode ($f_{3dB}$ ~37 GHz, Discovery Semiconductor). Wavelength separations below ~25 GHz are not resolvable as the corresponding Brillouin shifts become almost identical for both of the pumps, rendering both of them to fall into a singular wavelength operation via the self – injection scheme, effectively eliminating the beat note. A stable beat note with measureable phase noise PSD is recorded around 27.3 GHz (Fig. 2b). The phase noise PSD of the directly measured Δλ of 27.3 GHz features exceedingly low phase noise levels; roughly -98 dBc/Hz, -114 dBc/Hz and -120 dBc/Hz corresponding to Fourier frequencies of 1 kHz, 10 kHz and 100 kHz, respectively (Fig. 2b). In anticipation of facilitating higher frequencies aimed towards mmW operation, we move the operating wavelengths of the pump lasers further apart and adjust for the corresponding Brillouin shifts in each feedback channel to sustain the self–injection condition. In order to measure $2-\lambda$ frequencies in the mmW/THz regime and associated high frequency range without necessitating such high speed photodiodes and measurement equipment, we employ alternative techniques to facilitate effective measurement of $2-\lambda$ frequencies ranging in 100's of GHz (see supplementary information). We channel the $2-\lambda$ Stokes light into an electro–optic (EO) comb for multiple sideband generation and subsequently through an optical bandpass filter (OBPF) to filter out the central beat note based on the number of sidebands and the $\Delta\lambda$ frequency in question. This central beat note of the associated sidebands essentially contains a proxy for the $2-\lambda$ frequency beat in corresponding noise and stability properties while residing within a reduced frequency range, facilitating the usage of available laboratory equipment. The EO comb features three phase modulators connected in series and driven by a dielectric resonant oscillator (DRO) at around 10 GHz. The intermediate frequency of the central sideband beat note is then photo-detected and analyzed with a commercial phase noise analyzer (Agilent PXA N9030A). Note, one important drawback in this process is that the DRO noise is also added to the base noise of the $2-\lambda$ Stokes signal (Fig. 2b, see supplementary information). The maximum operating range of the pump lasers in use allow up to about 0.52 THz operation. We measured the phase noise PSD by varying the pump currents resulting in $\Delta\lambda$ frequencies ranging from 180 – 520 GHz. The phase noise of these $2-\lambda$ frequencies feature elevated levels compared to the directly measured 27.3 GHz signal, performing around -50 dBc/Hz, -80 dBc/Hz and -100 dBc/Hz levels at Fourier frequencies of 1 kHz, 10 kHz and 100 kHz, respectively (Fig. 2b). This is primarily due to the additive DRO noise from the employed measurement scheme as the individual SBS lasers should not impose elevated noise levels from just increasing the spacing between the two pump wavelengths. The fact that any of the two lasers can get noisier from nonlinearities imposed by higher current operation can also be ruled out as evidenced by the prior SBS tunability with wavelength results (Fig. 1c).

The $2-\lambda$ Stokes signal is also measured using a frequency counter to obtain fractional frequency stability (Fig. 2c). The modified Allan deviation, $\sigma_y^{Mod}(\tau)$ averages down to about 1.8×10$^{-10}$, 1.1×10$^{-10}$ and 7.3×10$^{-11}$ around 1 ms averaging time corresponding to $\Delta\lambda$ frequencies of 301 GHz, 382 GHz and 502 GHz; respectively. The fractional frequency stability improves as the wavelength difference increases. This result is not unexpected as the two pump lasers move further away and behave more independently towards their respective self – injection criterion. This trend in increasing the stability of the $\Delta\lambda$ frequencies with increasing the spacing between the pump lasers also indicates constant phase noise levels for associated increased $2-\lambda$ spacings while reducing fractionally.

## Sub-mmW Generation and Characterization

Consequently, we collect the $2-\lambda$ Stokes signal from the Brillouin cavity and beat on a uni-travelling carrier photodiode (UTC-PD) capable of producing sub-mmW signals. First, we select operating wavelengths of the pump CW lasers aimed at desired sub-THz separations and self–injection lock both the pump lasers based on aforementioned principles considering the operating region of the pump diodes and equipment availability. We utilize a 600 GHz UTC-PD coupled to a sub-mmW waveguide output [25]. We employ a particular setup for characterizing the noise properties of the $2-\lambda$ sub-mmW signal based on a self-heterodyne interferometer and a down-conversion mechanism based on a UTC-PD coupled to an amplitude detector with mmW waveguides (Fig. 3). A self-heterodyne interferometer is configured to receive the $2-\lambda$ signal corresponding to a sub-mmW frequency of 0.5 THz (Fig. 3a(i)). Two 80:20 fiber power splitters are used to direct the $2-\lambda$ Stokes light into separate channels and recombine again at the output of the interferometer. The higher power channel comprising the first arm of the interferometer is once again split off and recombined using a pair of 50:50 fiber couplers to construct a nested interferometer with two OBPFs on either sub-arm to filter out each of the SBS lasers from the $2-\lambda$ signal and an acousto-optic modulator (AOM1) in one sub-arm to translate one of the SBS laser frequency by the AOM drive frequency, $f_{AOM1}$ which effectively frequency-shifts the input frequency difference by that amount. The OBPF bandwidths and frequencies are chosen in accordance with the available optical power in each SBS line so that the higher power line passes through the AOM which incurs insertion losses and the resultant optical signal from the output is more balanced in power (See supplementary information). The second arm (i.e. the lower power channel of the 80:20 fiber couplers) of the interferometer is configured to frequency-shift the input optical signal with another AOM (AOM2) driven at $f_{AOM2}$ in order to not interfere with the wavelengths of the first arm, and comprises a fiber delay line (of about 500 m) configured to delay the frequency-shifted $2-\lambda$ optical signal by a delay, $\tau$. The output from the interferometer is then amplified with an EDFA and passed through a VOA for precise control of optical power incident on the UTC-PD. Essentially, the UTC-PD receives four optical lines from the interferometer as $v_{s1}(t)$, $[v_{s2} + f_{AOM1}](t)$, $[v_{s1} + f_{AOM2}](t-\tau)$ and $[v_{s2} + f_{AOM2}](t-\tau)$ which generate four disparate mmW beat notes as $[v_{s2} - v_{s1}](t) + f_{AOM1}(t)$, $[v_{s2} + f_{AOM2}](t-\tau) - v_{s1}(t)$, $[v_{s2} + f_{AOM1}](t) - [v_{s1} + f_{AOM2}](t-\tau)$ and $[v_{s2} - v_{s1}](t-\tau)$. The generated mmW signals at the UTC-PD inhere to the difference frequencies as $f_{mmW1} = [v_{s2} - v_{s1}](t-\tau) + f_{AOM2}(t-\tau)$ and $f_{mmW2} = [v_{s2} - v_{s1}](t)$. These notes beat on a zero bias Schottky barrier diode (SBD) as an mmW amplitude detector configured to receive the mmW signals (Virginia Diodes, WR1.5ZBD). Acting as a low pass filter, the detected beat note of interest is $[v_{s2} - v_{s1}](t-\tau) + f_{AOM2}(t-\tau) - [v_{s2} - v_{s1}](t)$ and the phase noise that $f_{AOM2}$ carries is then modulated by $[v_{s2} - v_{s1}](\tau)$ which is the phase noise of interest, $Sf_{\delta x}(\tau) = S(v_{s2} - v_{s1})(\tau) + Sf_{AOM2}(\tau)$. The RF signal detected at the output of the mmW amplitude detector therefore contains the phase noise of the millimeter-wave oscillator under test, i.e. the $2-\lambda$ Stokes signal.

The phase noise of the 0.5 THz mmW radiation is calibrated from the RF signal phase noise similarly as outlined in the SBS phase noise measurement assisted by the self-heterodyne scheme (see methods and supplementary information). The calibrated phase noise PSD of the mmW signal features roughly -17 dBc/Hz, -76 dBc/Hz and -98 dBc/Hz levels at Fourier frequencies of 1 kHz, 10 kHz and 100 kHz, respectively. Next, we repeat the same measurement for operation at a mmW frequency of 300 GHz using a 300 GHz J-band UTC-PD (IOD-PMJ-13001, NTT Electronics) and a zero-bias SBD operating in the desired frequency range (Virginia Diodes, WR3.4ZBD). As timing noise is an effective figure of merit to compare different oscillators with disparate frequencies, we calibrate the phase noise PSD to timing noise PSDs. The achieved phase noise levels for the sub-mmW 500 GHz signal correspond to exceedingly low timing noise levels featuring 65 fs.Hz$^{-½}$, 70 as.Hz$^{-½}$ and 5.3 as.Hz$^{-½}$ at Fourier frequencies of 1 kHz, 10 kHz and 100 kHz, respectively (Fig. 3b). The white phase noise floor reaches 3 as.Hz$^{-½}$ beyond Fourier frequencies of 100 kHz. A couple of peaks are observed at Fourier frequencies corresponding to the multiples of the null frequency of about 400 kHz arising from the delay fiber length of approximately 500 m. Note, the achieved timing noise levels are roughly similar as a 300 GHz wave in dissipative Kerr soliton based on a state-of-the-art SiN microresonator despite a higher carrier frequency and offering an off-the-shelf low cost solution for low-noise applications without necessitating expensive foundries or cleanroom environments [10]. Simultaneously, the measured 300 GHz mmW signal with the same optical interferometric setup (Fig. 3a(i)) revealed similar levels of timing noise below 10 kHz Fourier frequencies. Timing noise levels achieved for Fourier frequencies above 10 KHz, and consequently, the white phase noise floor for the 300 GHz mmW signal are considerably lower than the 500 GHz case. The measured phase noise PSD matches exactly phase noise PSD of the synthesizer driving AOM2 (Fig. 3a(i), see supplementary information) above 10 KHz Fourier frequency, ultimately limiting the measurement. It is worthwhile to mention here that while the 600 GHz UTC-PD features a relatively flat response over the frequency range measured here (near 0.5 THz), the mmW amplitude detector Schottky diode exhibits reduced responsivity (~850 V/W, as opposed to the peak responsivity of 1700 V/W at 700 GHz) near this region as approaching its rated cutoff range around 0.5 THz. This coupled with the low responsivity of merely 0.14 A/W of the 600 GHz UTC-PD (in comparison, the J-band UTC-PD responsivity is 0.22 A/W) points to the limitations of the sub-mmW measurements in signal to noise as facilitated by the mmW hardware.

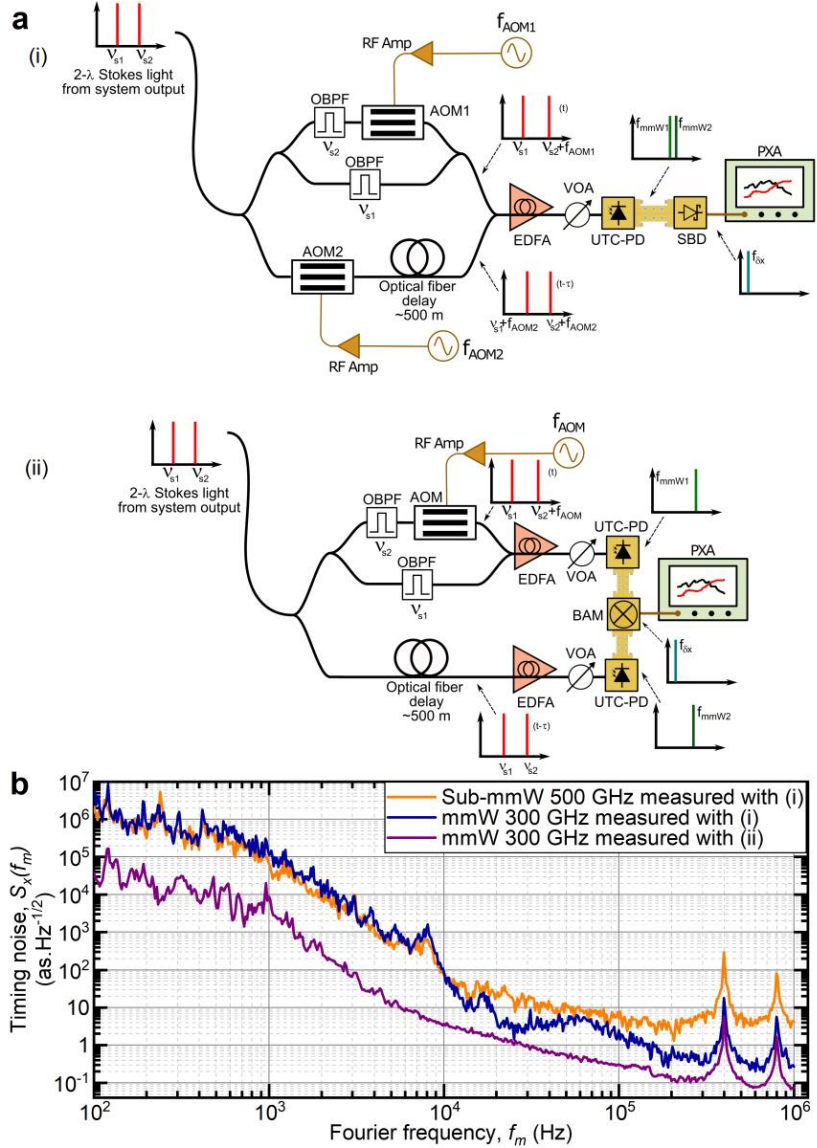

**Fig. 3: Sub-millimeter wave generation based on the two wavelength Stokes light from the self − injection scheme and associated phase noise characterization.** (**a**) Schematic illustration of sub-millimeter-wave phase noise characterization based on a self-heterodyne interferometer and a down-conversion mechanism with uni-travelling carrier photodiodes (UTC-PD) coupled to a millimeter-wave amplitude detector Schottky barrier diode (SBD) or a balanced amplitude mixer (BAM). (**i**) The two wavelength $(2 - \lambda)$ Stokes light from the Brillouin cavity is channeled into an interferometer consisting of a delay fiber line of ~500 m, two acousto-optic modulators (AOMs) – one of which AOM1 is within a nested sub-interferometer with two optical bandpass filters (OBPF). The output of the main interferometer is then opportunistically tuned with an Erbium doped fiber amplifier (EDFA) and a variable optical attenuator (VOA) to reach the UTC-PD which converts the optical frequency differences into millimeter waves that, in turn, beat on the SBD to produce an RF signal to analyze with a phase noise analyzer (PXA). (**ii**) The two wavelength $(2 - \lambda)$ Stokes light from the Brillouin cavity is channeled into an interferometer consisting of a delay fiber line of ~500 m, an AOM within a nested sub-interferometer with two OBPF. Both outputs from the nested interferometer and the delay line are tuned in power with two sets of EDFA and VOA to reach two UTC-PDs which convert the optical frequency differences into millimeter waves and frequency mixed with a BAM to produce an RF signal to analyze with a phase noise analyzer (PXA). (**b**) Calibrated timing noise, $S_x(f_m)$ (as.Hz$^{-\frac{1}{2}}$) power spectral densities (PSD) measured with the setup outlined in (i) for the sub-millimeter wave 0.5 THz signal (orange), millimeter wave

0.3 THz signal (blue) corresponding to Fourier frequencies, $f_m$ (Hz) ranging from 100 Hz to 1 MHz. Timing noise PSD for the 0.3 THz millimeter wave signal measured with the setup outlined in (ii) is also presented in purple.

In order to surpass the limitations to characterize noise properties of the 300 GHz signal, we employ an alternative setup facilitated by the availability of two J-band UTC-PDs and a mmW frequency mixer at our disposal (Fig. 3a(ii)). Similar to the previous setup (Fig. 3a(i)), we channelize the $2-\lambda$ Stokes light in two paths – one with a nested interferometer including an AOM on one sub-arm and two OBPF in either sub-arm, and a delay line of about 500 m on the other path. We then pass each channel through a set of EDFA and VOA for precise optical power control and beat each channel on different UTC-PDs. The first channel (upper arm in Fig. 3a(ii)) carries two optical lines incident on the respective UTC-PD as $\nu_{s1}(t)$ and $[\nu_{s2}+f_{AOM1}](t)$ which generates a mmW beat note of $f_{mmW1} = [\nu_{s2}-\nu_{s1}](t)+f_{AOM}(t)$. The second channel (lower arm in Fig. 3a(ii)) carries the delayed $2-\lambda$ lines to the other UTC-PD generating a mmW beat note as $f_{mmW2} = [\nu_{s2}-\nu_{s1}](t-\tau)$. These two mmW frequencies are then mixed together in the mmW domain with a balanced mixer (Virginia Diodes, WR3.4BAM) operating at the desired frequency range of J-band. The difference in this setup from the previous setup is the mmW frequency mixing and reduction of an AOM in this setup. Phase noise levels achieved through this mmW frequency mixing for the 300 GHz signal correspond to exceedingly low timing noise levels featuring 8 fs.Hz$^{-½}$, 3.7 as.Hz$^{-½}$ and 0.3 as.Hz$^{-½}$ at Fourier frequencies of 1 kHz, 10 kHz and 100 kHz, respectively (Fig. 3b). The white phase noise floor reaches a miniscule 70 zs.Hz$^{-½}$ beyond Fourier frequencies of 100 kHz. Null frequency peaks at multiples of 400 kHz arising from the delay fiber length of approximately 500 m can be seen. A low noise frequency synthesizer (Agilent N5181A) along with a frequency divider (Valon 3010a) is used in this measurement to drive the AOM, but the phase noise is still limited by the source beyond 100 kHz Fourier frequencies (See supplementary information), suggesting the white phase noise floor of the mmW signal can actually lie beneath the measured value which is beyond the scope of measurement with the available equipment.

## Conclusion

We have demonstrated a feasible option for enhancing the spectral purity of inexpensive semiconductor diode lasers featuring mediocre performance (e.g. large linewidth ~800 kHz, side mode oscillations degrading SNR ~40 dB, heightened phase noise levels incapable of measurement with sophisticated lab equipment, etc.) via self–injection locking with corresponding stimulated Brillouin scattered Stokes waves simultaneously from a single Brillouin cavity in order to facilitate low noise millimeter to terahertz waves. Our results showcase significant spectral purity of output SBS laser beam over the pump CW laser featuring sub-100 Hz linewidth, greatly reduced phase noise levels, and tunable operation with suppressed phase noise levels approximately -53 dBc/Hz, -85 dBc/Hz and -110 dBc/Hz levels at Fourier frequencies of 1 kHz, 10 kHz and 100 kHz, respectively. Further demonstration of difference frequency generation in the sub-millimeter wave region revealed promising results in terms of noise and stability characteristics bolstering miniscule levels of phase noise. Finally, we have experimentally demonstrated the first photonic sub-millimeter 0.5 THz wave oscillator offering zeptosecond level timing noise. It should be noted that all such improvements are realized utilizing only off-the-shelf readily available components. Such refinement of cheap sub-optimal pump lasers enables involvement of inexpensive options without necessitating the value to performance tradeoff aimed at a plethora of low noise applications including micro- or millimeter-wave photonics, spectroscopic applications, LiDAR applications in velocimetry, etc.; for instance. The entire setup is quite modest in size, conforming to a lab-scale footprint on couple of optical breadboards benefitting the added advantage of rack-mounted portability. Recent advances on integrated photonics and possibilities looming over the horizon can enable a chip-scale integration of this self–injection scheme with micro-ring based Brillouin cavities and waveguide based EOMs [26-30].

## Methods

### Self – injection of a diode laser based on stimulated Brillouin scattering

We start off with a relatively large linewidth semiconductor diode laser (~800 kHz, reported in corresponding manufacturer datasheet) incorporating no optical isolator (Eblana Photonics EP1550-0-DM-B05-FM, DCP2644), operating in the telecom C – band near infrared (NIR) region. The CW laser output is passed through a 90:10 fiber power splitter (Fig, S2, see supplementary information). The greater optical power output is then fed into an Erbium doped fiber amplifier (EDFA), the output of which subsequently non-resonantly pumps a 75 m long fiber Brillouin ring resonator. The corresponding acoustic damping from the silica fiber redshifts the light output by the amount of the resultant Brillouin shift, $\nu_B$ at the operating wavelength, $\lambda$. The Stokes output is next passed through an electro-optic modulator (EOM) in order to transfer the Brillouin shifted stokes light back into the frequency of the pump depending upon the operational frequency of the EOM which is set to match the Brillouin shift, $\nu_B$. We drive the EOM with a frequency synthesizer (HP 8341A) followed by a high power RF amplifier operated at 1 W in the desired frequency range. We use a Mach-Zehnder intensity modulator for the EOM in our setup. The output from the EOM is then passed through variable optical attenuator (VOA) for precise control of the optical power, which is subsequently injected back into the CW laser via the lower power channel of the initial fiber coupler. The resultant output from the EOM contains sidebands generated from operation which can be expressed in frequencies as $\nu_{LD} \pm mf_{EOM} \equiv \nu_{LD} \pm m\nu_B$; where $\nu_{LD}$ is the CW output frequency of the pump, and $m$ is an integer corresponding to the order of the EOM sidebands. In our operation, we only require the first blue-shifted sideband from modulation, $m = -1$ to be injected back into the pump laser; the remaining sidebands from EOM operation do not need to be filtered out as the pump frequency, $\nu_{LD}$ is not affected by those. As such, the diode laser can be self – injection locked and improve its output efficiency iteratively in the loop until a thermal equilibrium of the Brillouin cavity noise floor is reached. The Stokes radiation from the Brillouin cavity is the crucial output of our system manifesting an SBS laser.

### Phase noise characterization of a self – injected SBS laser

We investigate the phase noise of the SBS output independently, i.e. the out of loop signal, with a delayed self-heterodyne measurement setup (Fig. S3). The self-heterodyne measurement features a Mach-Zehnder interferometer (MZI) with an acousto-optic modulator (AOM) on one arm and a delay line on the other (see supplementary information). It effectively positions the input radiation at the AOM frequency, $f_{AOM} = 80\ MHz$ for measurement ease and equipment compatibility. The phase noise of the SBS laser is calibrated from the RF signal phase noise as $\pounds_{SBS}(f_m) = \pounds_{RF}(f_m) - 20\log(2\sin(\pi f_m \tau_d))$ [31], where $\pounds_{RF}$ is the RF signal phase noise, $f_m$ is the Fourier frequency from the carrier, and $\tau_d = nL_d/c$ is the time delay on one arm of the MZI. Here, $n$ is the effective refractive index of the single mode fiber (SMF), $c$ the celerity of light in vacuum, and $L_d$ is the delay fiber length. The phase noise PSD of the Stokes output features roughly -53 dBc/Hz, -85 dBc/Hz and -110 dBc/Hz levels at Fourier frequencies of 1 kHz, 10 kHz and 100 kHz, respectively (Fig. S4a). Interestingly, the SBS laser linewidth can also be estimated from the phase noise measurements from the transition of the polarity of the phase noise, i.e. at the Fourier frequency where the phase noise PSD crosses zero. Our measurements estimate an SBS laser linewidth of roughly 40 Hz from this crossover Fourier frequency, which is a remarkable ~2,000× reduction of the reported pump linewidth resulting from the demonstrated injection locking scheme. It is worthwhile to mention here, the intrinsic natural linewidth deduced from the $\beta$-separation line and the SBS laser frequency noise floor is approximately 44 mHz (see supplementary information) [32, 33].

### Fractional frequency stability and relative intensity noise characterization of a self – injected SBS laser

The fractional frequency stability of the self – injection scheme for a single laser diode was measured by counting the frequency of the beat note between the Stokes wave and the pump laser (which is, in fact, equivalent to an in-loop error). The frequency counter was operating in $\lambda$-mode at 1 second gate time and clocked by an Rb reference at 10 MHz [34]. It exhibits an Allan deviation, $\sigma_y(\tau)$ of approximately 1.8×10$^{-16}$ at 1 second averaging time (Fig. S4b). The modified Allan deviation, $\sigma_y^{Mod}(\tau)$ averages down to about 5×10$^{-19}$ over 100 s averaging time, indicating stable operation of the self – injection scheme. This means that further isolation and/or passive stabilization of the cavity would further reduce the linewidth of the pump down to that residual instability reported here.

Additionally, the relative intensity noise (RIN) for the output SBS laser and the self – injection locked pump laser are characterized using a low noise photodetector (Thorlabs, PDA05CF2) and subsequent spectrum analyzer (see supplementary information). The RIN measurement involved PSD calibration of Fourier frequencies ranging between 10 kHz and 10 MHz by measuring the associated RIN within each decade of the Fourier frequencies and then stitching them together (Fig. S4c). A DC block stopping frequencies below 10 kHz is used in these measurements to refrain from damaging the spectrum analyzer. The RIN PSD associated with the Stokes shifted output SBS laser features nearly -115 dB/Hz levels at 10 kHz and reduces down to about -156 dB/Hz towards higher Fourier frequencies (> 1 MHz). The SBS laser RIN PSD exhibits distinct peaks at multiples of ∼ 2.5 MHz corresponding to the free spectral range (FSR) of the Brillouin resonator and possibly owing to a small amount of Stokes light being generated at other cavity resonances in operation (Fig. S4c). The pump laser under a self – injection locked condition is also investigated and reveals elevated RIN performances at Fourier frequency of 10 kHz featuring about 127 dB/Hz while reducing down to about -158 dB/Hz at Fourier frequencies > 1 MHz (Fig. S4c). Interestingly, the locked pump laser PSD does not feature any distinct cavity resonance peaks from the Brillouin resonator despite being fed back by the Stokes wave. This reflects the narrow bandwidth of the self – injection scheme in the Fourier domain benefitting particularly the central region of the laser beam in Fourier frequency space. Also, the decreased RIN at lower Fourier frequencies in the pump PSD compared to the SBS denotes the stability augmentation of the pump cavity which is considerably shorter, and therefore, less susceptible to minute mechanical vibrations and pressure fluctuations, unlike the Brillouin cavity. Note, the self – injected pump can similarly serve as an output to the self –injection system demonstrated herein depending upon application specifications, which might potentially offer distinctive benefits in operations where intensity fluctuations are crucial. A dark current measurement is also conducted by disconnecting the fiber input to the photodiode and performing similar evaluation to gauge the electronic noise floor (approximately -165 dB/Hz) of these measurements (Fig. S4c).

# Supplementary Information

# Exceeding octave tunable Terahertz waves with zepto–second level timing noise

Rubab Amin[1], James Greenberg[1], Brendan Heffernan[1], Tadao Nagatsuma[2], and Antoine Rolland[1]

[1]IMRA America, Inc., Boulder Research Lab, 1551 S. Sunset St. Suite C, Longmont, Colorado 80501, USA

[2]Graduate School of Engineering Science, Osaka University, Osaka 560-8531, Japan

## 1. Brillouin Cavity Threshold

A 75 m long fiber Brillouin cavity used to generate the Stokes shifted stimulated Brillouin scattered (SBS) laser is placed inside a vacuum chamber coupled with temperature control and mechanical stabilization. The temperature was kept constant at 28ºC within the enclosure. Mechanical vibration isolation for the cavity is achieved by placing the cavity in a vacuum chamber atop a bench top vibration isolation platform (Minus K negative stiffness vibration isolator, 50BM-10) on an optical table. The optical Table (Thorlabs PTM11109) features its own passive vertical isolation support (Thorlabs PTP603). The vacuum pump was monitored using a convection enhanced Pirani vacuum gauge module with a built-in controller, display and process control signal (Stinger by InstruTech, Inc.; CVM211-

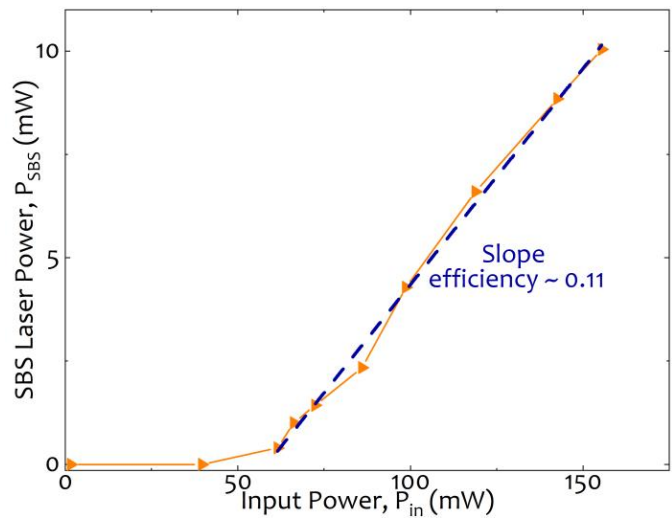

**Fig. S1: SBS Laser threshold.** The Brillouin shifted stokes laser features a threshold optical power of approximately 60 mW. The SBS laser operates with a slope efficiency of ~ 0.11 beyond threshold.

GCL). The Brillouin cavity is constructed using polarization maintaining (PM) fibers and an optical fiber circulator such that only the Brillouin shifted Stokes light is resonant on the cavity and the pump light is not; i.e. the Brillouin cavity features non-resonant pumping condition.

The Brillouin cavity is pumped with the pump laser light amplified using a two stage erbium doped fiber amplifier (EDFA). The EDFA is an in-house constructed ad-hoc module featuring two pump lasers operating at 980 nm and portions of erbium (Eb) doped fibers. The Brillouin cavity threshold is found to be approximately 60 mW of optical input power (Fig. S1). The SBS laser exhibits a characteristic slope efficiency of nearly 0.11 (Fig. S1). Note, this efficiency is inherently quite low due to the non-resonant pumping condition, similar fiber cavities with resonant pumping usually feature higher slope efficiency. In all our experiments, the Brillouin cavity was always pumped above the corresponding threshold.

## 2. Experimental Setup for the Self–Injection scheme

As outlined in methods, we tap off a portion of the pump power right after the EDFA with another 90:10 power splitter where the higher power output is sent to the Brillouin cavity; we again tap off a portion of the Stokes radiation from the Brillouin cavity output with a 50:50 power splitter and beat both the Stokes and pump (i.e. the lower optical power output of the first 90:10 coupler) signals on a fast photodiode (~18 GHz) and analyze the signal on an electrical spectrum analyzer utilizing another 2×2 50:50 fiber power splitter. The frequency of the beat note (i.e. the in-loop signal) corresponds to the Brillouin shift in the fiber (Fig. S2b). The other output of the 2×2 splitter is monitored with an optical spectrum analyzer (OSA) and the corresponding peaks for the pump and Stokes light can be observed distinctly when the injection scheme is not functional (Fig. S2c).

We engage the self – injection scheme by turning the frequency synthesizer on with the desired frequency corresponding to the Brillouin shift in fiber, $\nu_B$. Categorically, the iterative operation, where the fed back portion of the Stokes refines the pump by copying the concurrent phase noise properties of the Brillouin cavity into the pump and the refined pump further refining the Stokes and so on, until a thermal threshold of the cavity is reached; produces an intense and spectrally narrow beat on the photodiode which is resolvable down to 1 Hz resolution levels. The self – injection scheme goes out of lock, i.e. turned off, when we disengage the frequency synthesizer. While both the

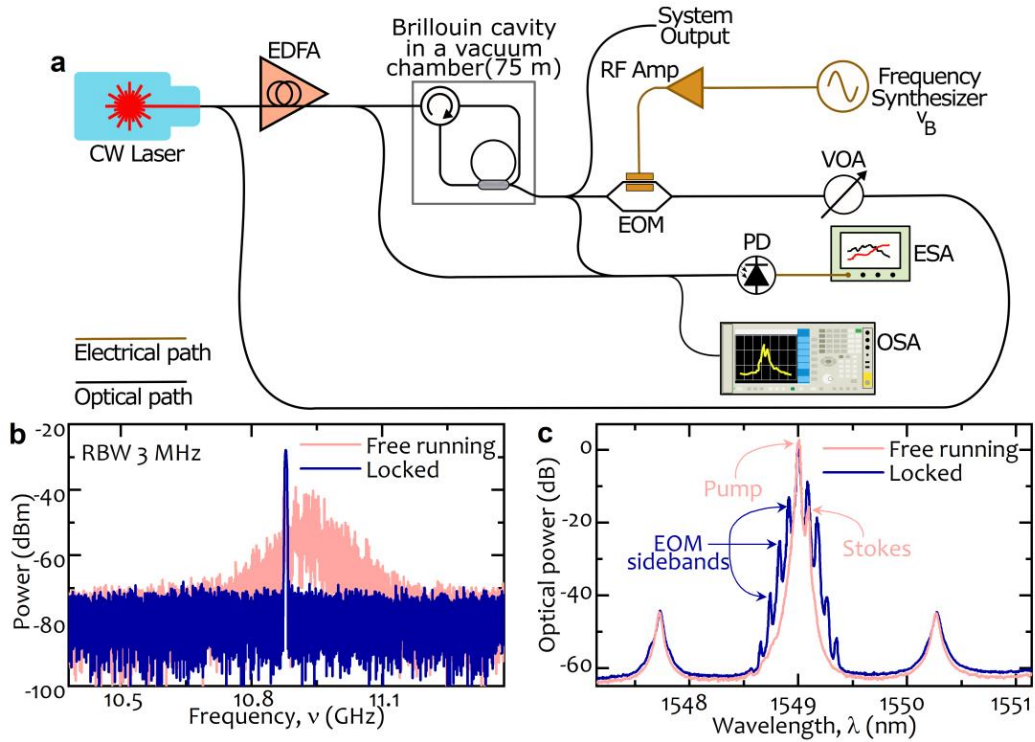

**Fig. S2: Self – injection locking setup and the pump – Stokes beat note. (a)** A CW pump laser is amplified with an Erbium doped fiber amplifier (EDFA) and sent to a non-resonantly pumped Brillouin cavity. The backwards propagating Stokes light is collected from the cavity and passed through an electro-optic modulator (EOM) driven at the Brillouin shift frequency in fiber, $\nu_B$ with the help of a frequency synthesizer and RF amplifier. The EOM generates sidebands according to the synthesizer frequency, which is subsequently fed back into the pump laser via a variable optical attenuator (VOA) for controlling the feedback power. Pump and Stokes light is tapped off to beat on a fast photodiode (PD) and observed with an electrical signal analyzer (ESA) and optical signal analyzer (OSA) simultaneously. The Stokes radiation from the Brillouin cavity is the output of the system. **(b)** Pump – Stokes beat note on the PD corresponding to the Brillouin shift of 10.879 GHz in free running and locked conditions. Resolution bandwidth (RBW) is 3 MHz in both cases. **(c)** Pump –Stokes beat optical spectra in the free running and locked conditions showing respective peaks of the pump and Stokes light along with EOM sidebands from operation in the locked state.

Stokes and pump light still remain there, which can be easily confirmed by the OSA, the rather large linewidth and mediocre noise performance of the used inexpensive pump laser forces the pump-Stokes beat to be considerably broader and weaker compared to the injection locked state (Fig. S2b). The broad full width half maximum (FWHM) spectrum of the unlocked beat indicates the large linewidth of the pump (and of course, the associated Stokes which is broader than the locked state), i.e. the broad linewidth of the pump results in broadening of the beat note. The near-MHz level linewidth of the pump forces the beat to be such broadened as to necessitating a very large span ~1 GHz on the spectrum analyzer to clearly visualize the beat note (Fig. S2b).

## 3. Self – Heterodyne Setup

The self – heterodyne setup for measurements of the SBS laser phase noise power spectral densities featured a fiber based Mach – Zehnder interferometer with an acousto-optic modulator (AOM) on one arm and a fiber delay line of 87 m on the other arm (Fig. S3). The AOM is used as a translational frequency component for measurement ease and compatibility of available measurement equipment as the AOM shifts the frequency by 80 MHz for simple electronic detection. Note, a delay line of 87 m is sufficiently greater than the inverse coherence time of the pump CW laser.

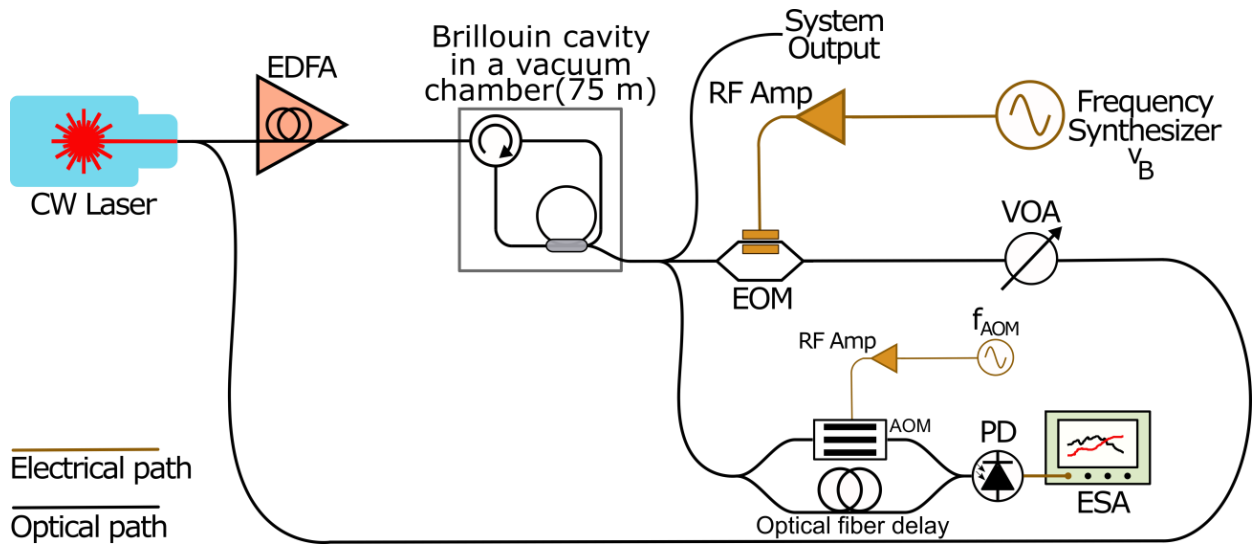

**Fig. S3: Self-heterodyne measurement setup within the self – injection scheme.** The Stokes output of the Brillouin cavity manifested by the stimulated Brillouin scattered (SBS) laser is the main output of the self – injection system. A Mach – Zehnder interferometer with an acousto-optic modulator (AOM) on one arm and a fiber delay line on the other arm establishes the delayed self-heterodyne setup for measurements of phase noise of the SBS laser.

## 4. Self–Injection Results and Discussion

The measured phase noise of the in-loop signal, i.e. the pump-stokes beat, shows reduced phase noise levels with approximately -93 dBc/Hz, -95 dBc/Hz, and -111 dBc/Hz levels at 1 kHz, 10 kHz and 100 kHz Fourier frequencies from the carrier, respectively (Fig. S4a). The in-loop phase noise power spectral density (PSD) follows the frequency synthesizer noise closely, indicating the constraint of the measurement system to be limited by the frequency synthesizer. Phase noise measurements are accomplished using a commercial phase noise analyzer (Agilent PXA N9030A).

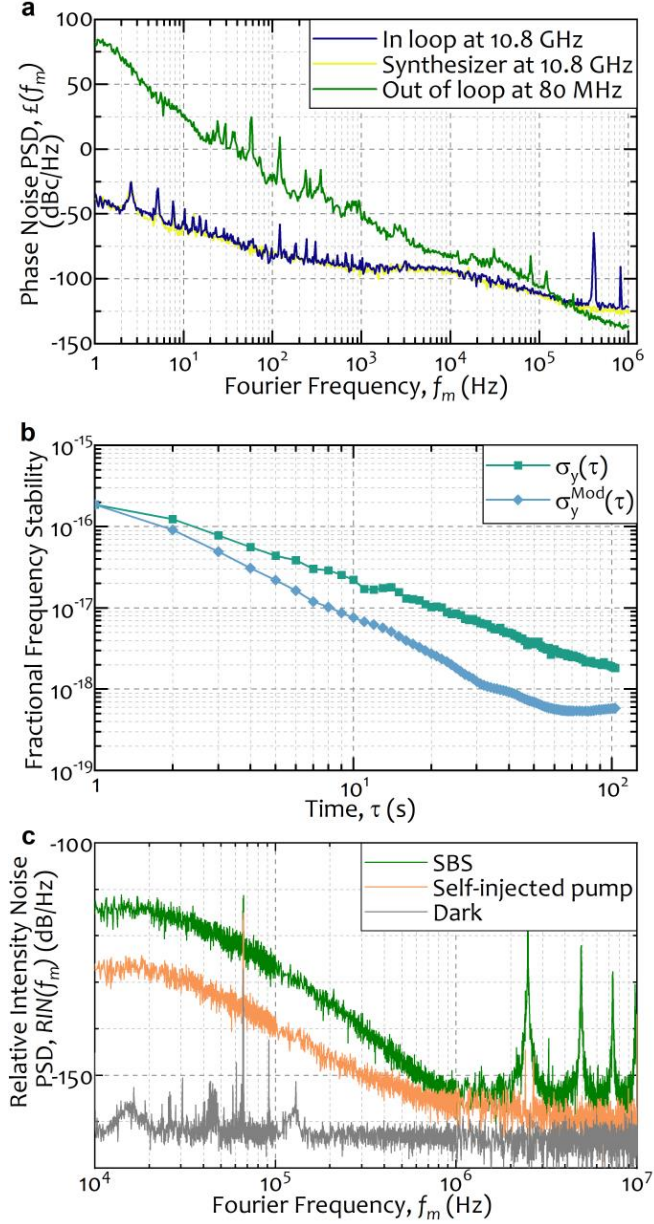

**Fig. S4: Noise properties attained with self –injection locking. (a)** Phase noise power spectral density (PSD), $£(f_m)$ in dBc/Hz vs. Fourier frequency, $f_m$ in Hz for the in loop signal comprising of the pump – Stokes beat and out of loop signal featuring the Stokes output; ranging from 1 Hz to 1 MHz. The frequency synthesizer phase noise PSD is also presented which manifests the limitation of the in loop measurement. **(b)** In loop fractional frequency stability in terms of Allan deviation, $\sigma_y(\tau)$ and modified Allan deviation, $\sigma_y^{Mod}(\tau)$ at 10.8 GHz. (c) Relative intensity noise PSD, $RIN(f_m)$ in dB/Hz vs. Fourier frequency, $f_m$ in Hz for the output SBS laser and the injected pump laser. A dark current measurement shows the noise floor for the implemented measurement setup.

Nevertheless, we take the output of our self-injection locking scheme as the Stokes output from the Brillouin cavity formulating an SBS laser radiation as the out of loop signal (Fig. S5a). CW output from the pump laser features many side mode oscillations as a characteristic drawback of the used inexpensive diode laser featuring a rather large linewidth. The pump laser side mode suppression ratio (SMSR) is approximately 40 dB. The iterative operation through repeated cycles of the self – injection locking scheme refines the linewidth of the SBS laser owing to the finesse of the Brillouin cavity. The SBS linewidth, $\Delta v_s$ can be narrowed down compared to the pump linewidth, $\Delta v_p$ as $\Delta v_s = \Delta v_p/K^2$, where $K = 1 + \gamma_A/\Gamma_c$ [1]. Here, $\gamma_A = \pi \Delta v_B$ represents the damping rate of the acoustic wave, where $\Delta v_B$ is the FWHM of the Brillouin gain curve, and $\Gamma_c = -c \ln R /nL$ is the loss rate of the cavity; $c/n$ is the celerity of light in the fiber of length $L$ and $R$ is an amplitude feedback parameter [1]. The operation of the self – injection scheme entails iterative refining of the pump laser linewidth which further refines the associated stokes, i.e.

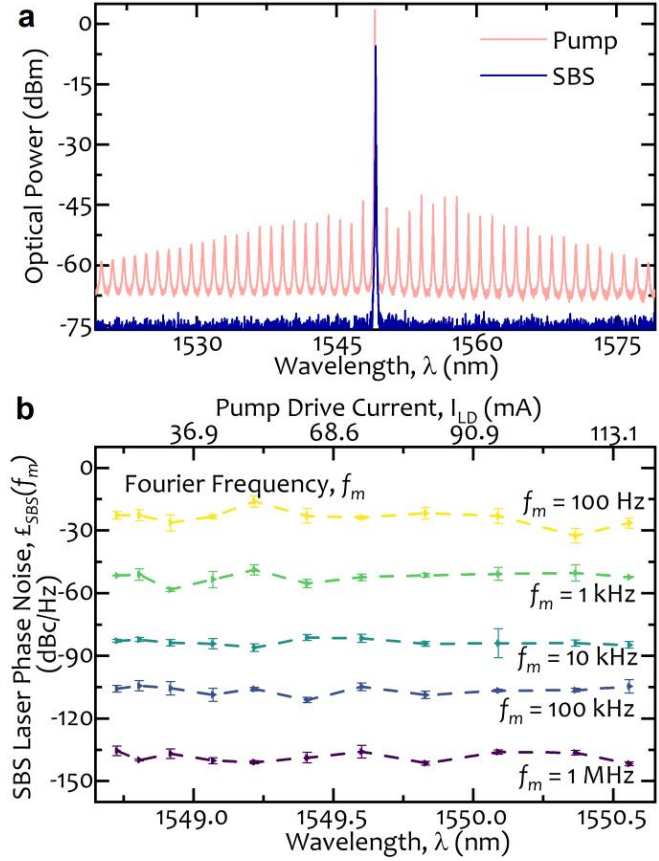

**Fig. S5: Performance characteristics of the out of loop stimulated Brillouin scattered (SBS) stokes output. (a)** Optical output spectra for the pump and SBS lasers. The pump laser exhibits a lot of side mode oscillations in non-resonant wavelengths and retains a side mode suppression ratio of about 40 dB, whereas the SBS output offers a highly clean spectrum with a single intense line at only the Brillouin shifted frequency with over 70 dB signal to noise ratio (SNR).

SBS linewidth, which again effects the pump linewidth and so on. This iterative refinements are persistent until a thermal threshold for the Brillouin cavity is reached. In the process of such refinements of the pump and SBS lasers within the self – injection scheme, the side mode oscillations from the pump vanish in the out of loop signal optical spectra offering a cleaner laser beam compared to the pump (Fig. S5a). The optical spectra confirms an SBS laser line with >70 dB intensity distinction from the noise floor of the OSA, which is >30 dB improvement over the pump in signal to noise ratio (SNR); achieved through the self – injection scheme.

## 5. SBS Laser Wavelength Tunability

The SBS laser wavelength tunability is a feature of the robustness of the self – injection scheme, which is obviously a byproduct of the pump laser wavelength tunability as a function of its drive current. A number of phase noise measurements were carried out to characterize the stability of the SBS laser with wavelength (also pump laser drive current, $I_{LD}$) tuning. It is worthwhile to mention here, the SBS laser output is evidently the output of the self – injection scheme. A set of full phase noise measurements for the entire tuning range is presented here (Fig. S6) to supplement the results shown in Fig. 1(c) in the main text.

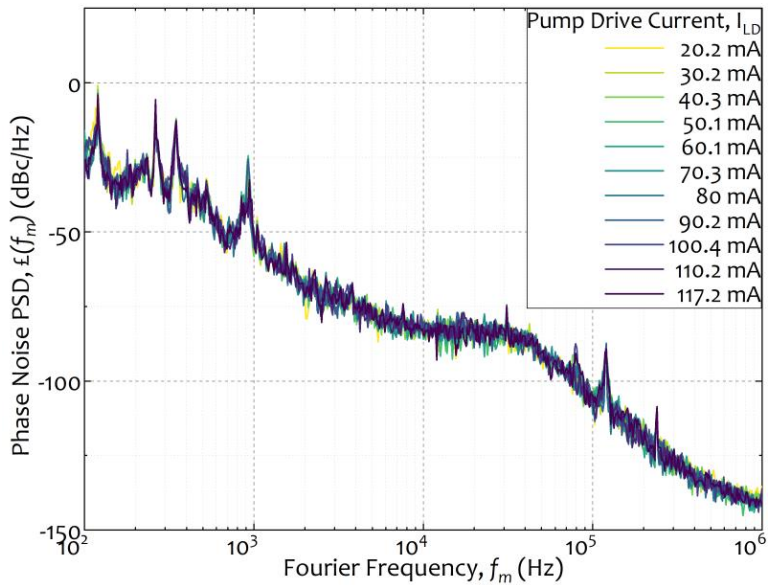

**Fig. S6: SBS Laser phase noise with wavelength tuning.** The pump laser diode is wavelength tunable with variations in its drive current, $I_{LD}$. Phase noise power spectral densities (PSDs) were evaluated at different operating wavelengths (corresponding to different pump laser drive currents, $I_{LD}$) for Fourier frequencies, $f_m$ ranging from 100 Hz to 1 MHz. The pump laser threshold current is 14 mA, and a maximum rated bias current of 120 mA is reported by the manufacturer. Accordingly we tuned the pump current in approximately 10 mA steps from 20 mA to 117 mA in corresponding experiments.

## 6. Pump – Stokes Beat with Wavelength Tuning

In order to characterize the wavelength tunability of the pump laser on the resultant SBS laser, we first chose to investigate the beat note between the pump and Stokes waves. As the Brillouin shift is a function of the operating wavelength, the modulation frequency needed to be altered with varying operating wavelengths from the current tuning of the pump laser. We tuned the pump laser in 10 mA drive current, $I_{LD}$ increments in the tolerable range while simultaneously optimizing the power and frequency of the synthesizer in each step. Our experiments revealed a rather wide range of tunable wavelengths from 1548.726 nm to 1550.579 nm corresponding to Brillouin shifts ranging from 10.8797 GHz to 10.868 GHz. The Brillouin shift is inversely proportional to the operating wavelength; as such, higher wavelength operation (i.e. higher pump drive current, $I_{LD}$) forces the Brillouin shift to decrease. Accordingly, the synthesizer frequency needed to be decreased gradually with tuning as well to match the Brillouin shift in the cavity, and hence, obtain a well-stabilized and strong beat signal (Fig. S7a). Phase noise measurements were carried out at each step by beating the pump – Stokes signal on a fast photodiode and analyzing with a phase noise analyzer (Agilent PXA N9030A). Phase noise power spectral densities (PSDs) were recorded accordingly at each step corresponding to the wavelength tuning within a Fourier frequency range from 100 Hz to 1 MHz (Fig. S7b). Phase noise levels corresponding to notable Fourier frequencies, $f_m$ with wavelength tuning are shown in Fig. S7(c). These results are directly comparable to the results in Fig. 1(c) in the main text where only the SBS laser phase noise levels corresponding to tuning are plotted. It is evident that the pump – Stokes beat phase noise levels are clearly vacillating and feature no useful trend with tuning whereas the SBS laser phase noise levels exhibit rather monotonous and near-constant levels. This could be due to the pump laser becoming progressively noisier with increasing drive current operation, i.e. beyond threshold current operation the nonlinearities in the gain medium of the laser feature progressively more noise as the output power increases exponentially. This interesting artifact points to the quality of the SBS laser's robust performance as, even with the wavelength tuning, it can sustain unvarying levels of phase noise performance for the chosen Fourier frequencies.

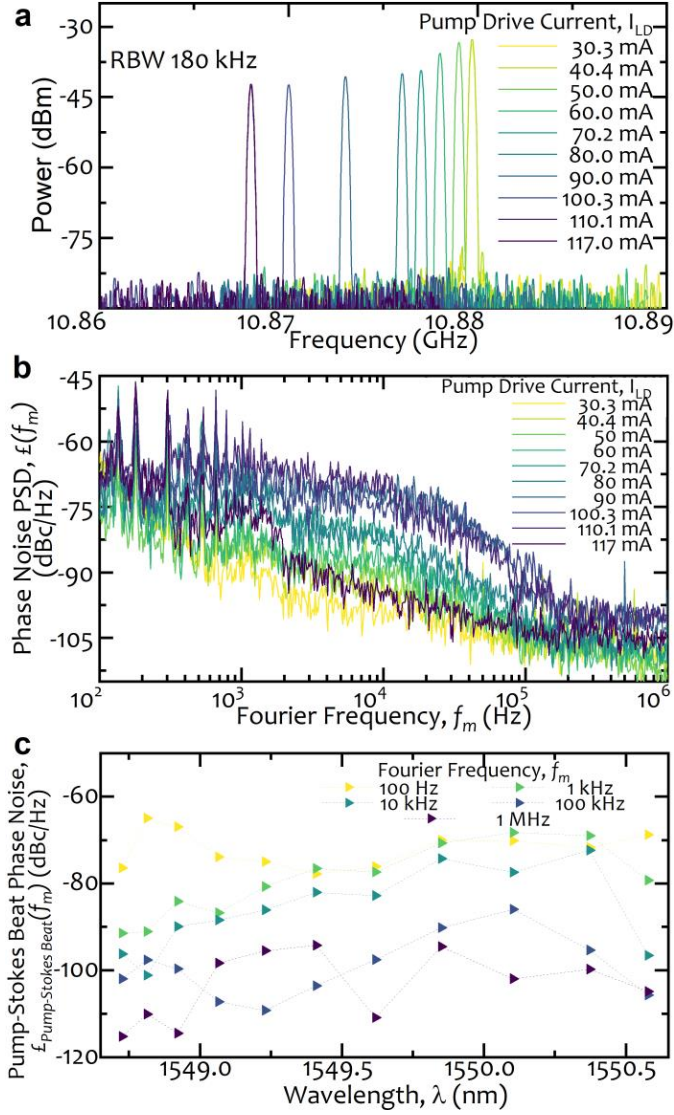

**Fig. S7: Pump – Stokes beat tunability characterization. (a)** The pump – Stokes beat with changing operation wavelength. The synthesizer frequency is also varied simultaneously to achieve the strong and narrow beat resonance. The resolution bandwidth (RBW) for all measurements is kept at 180 kHz. **(b)** Phase noise power spectral densities (PSDs) were evaluated at different operating wavelengths (corresponding to different pump laser drive currents, $I_{LD}$) for Fourier frequencies, $f_m$ ranging from 100 Hz to 1 MHz. **(c)** Pump – Stokes beat phase noise levels, $L_{Pump-Stokes\ Beat}(f_m)$ in dBc/Hz with wavelength tuning for Fourier frequencies, $f_m$ of interest i.e. 100 Hz, 1 kHz, 10 kHz, 100 kHz, and 1 MHz.

## 7. Brillouin Shift with Operating Wavelength

For pure backward Brillouin scattering in fibers, the Brillouin shift can be calculated from the effective refractive index $n_{eff}$, the acoustic velocity $v_a$, and the vacuum wavelength $\lambda$ as [2]

$$\nu_B = \frac{2n_{eff}v_a}{\lambda} \quad (1)$$

The Brillouin frequency shift depends on the material composition and to some extent the temperature and pressure of the medium with such dependencies exploited for fiber-optic sensors [2]. Brillouin scattering occurs only in a quite limited bandwidth, e.g. of the order of 100 MHz (which is a very tiny fraction of the optical frequency) in the case of silica fibers [2]. It depends on the damping time of the involved acoustic wave (the phonon lifetime), but can also be inhomogeneously increased, e.g. when the temperature of the active fiber in a fiber amplifier varies along the length [2].

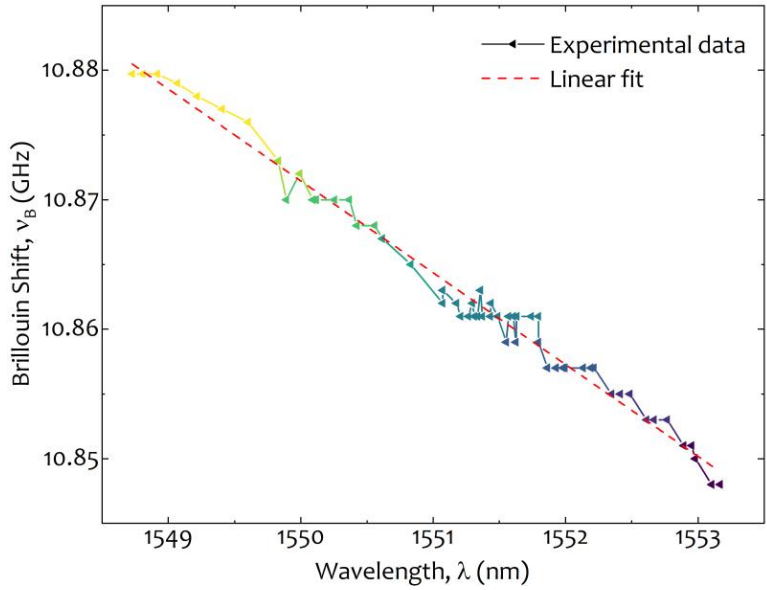

**Fig. S8: Brillouin shift in fiber, $\nu_B$ (GHz) corresponding to the operating wavelength, $\lambda$ (nm) of the stimulated Brillouin lasers.** The pump – Stokes beat with changing operation wavelength is measured to ascertain the Brillouin shift in fiber. The synthesizer frequency is varied simultaneously to achieve the strong and narrow beat resonance.

We find the Brillouin shift in our used fibers utilizing the pump – Stokes beat by varying the operating wavelength and adjusting synthesizer frequency in order to facilitate a strong beat note corresponding to the Brillouin shift frequency on the fast photodiode (Fig. S8). We utilize the full tuning range of the pump lasers centered around 1500 and 1552 nm which correspond to about 1.8 nm. The Brillouin shift frequency in fiber that we find with this self – injection locking method aligns closely to reported values in the vicinity of ~10.8 GHz. The strong pump –Stokes beat can be observed within a singular locking frequency with about 5 MHz of locking range which corresponds to the locking bandwidth for each Brillouin frequency shift bandwidth. A linear fit as indicated by eq. (1) reveals a decreasing slope of -0.0071 GHz/nm with a y-intercept of about 21.87 GHz with a coefficient of determination for the linear fit, $R^2 = 0.99$.

## 8. SBS Frequency Noise

The SBS laser frequency noise power spectral density, $S_{\delta\nu}(f_m)$ is calculated from the phase noise power spectral density of the self – heterodyne signal at 80 MHz for Fourier frequencies ranging from 10 Hz to 1 MHz (Fig. S9). The straight orange dashed line accounts for the $\beta$ – separation line [3]. The $\beta$ – separation line, given by $S_{\delta\nu}(f_m) = 8\ln(2)\, f_m/\pi^2$ [3], separates the SBS laser frequency noise spectrum into two regions whose contributions to the SBS laser line shape is quite different. The high modulation index area is the area under the frequency noise curve from the intersection point of the $\beta$ – separation line and the frequency noise spectra to the left, and contributes to the linewidth; whereas the low modulation index area is the area under the frequency noise curve from the intersection point of the $\beta$ – separation line and the frequency noise spectra to the right, and contributes only to the wings of the line shape. The $\beta$ – separation line reveals a flicker noise linewidth of approximately 141 Hz for an integration time of 0.1 s. Additionally, the flattening of the frequency noise power spectral density down to a white-noise level, $N_w$ is evident from Fig. S9, and yields an intrinsic natural linewidth, $\delta v_{exp} = \pi N_w$ of nearly 44 mHz [4].

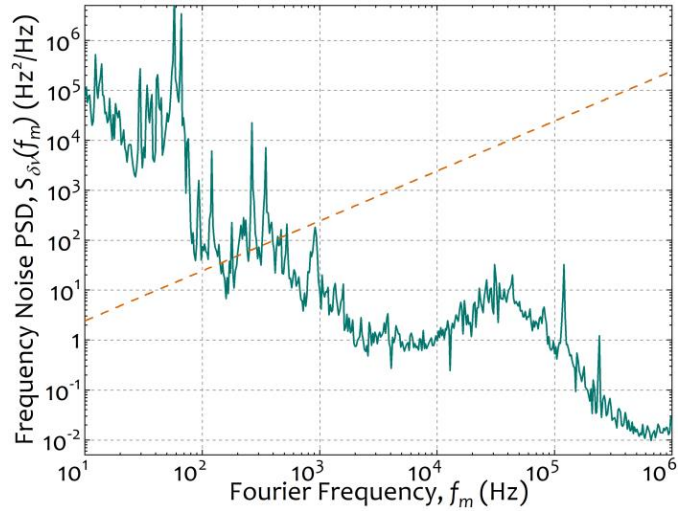

**Fig. S9: SBS laser frequency noise power spectral density.** Frequency noise, $S_{\delta\nu}(f_m)$ is calculated from the phase noise measurements for Fourier frequencies ranging from 10 Hz to 1 MHz. The orange straight dashed line depicts the $\beta$ – separation line [3].

## 9. RIN Measurements

The relative intensity noise (RIN) of a laser is a measure of the power fluctuations of a laser as $P(t) = \bar{P} + \delta P(t)$ [2]. Formally, the intensity noise can then be described statistically with a PSD of $S_l(f_m) = \frac{2}{\bar{P}} \int_{-\infty}^{\infty} \langle \delta P(t) \delta P(T + \tau) \rangle \exp(i2\pi f_m \tau) \, d\tau$ [2]; since $V = \eta R P$, we note that $S_l(f_m)$ can be rewritten as

$$S_l(f_m) = \frac{2}{\bar{V}^2} \int_{-\infty}^{\infty} \langle \delta V(t) \delta V(T + \tau) \rangle \exp(i2\pi f_m \tau) \, d\tau \tag{2}$$

where $V$ is the DC voltage at the output of the transimpedance amplifier, $\eta$ and $R$ are the quantum efficiency and TIA gain, respectively. The result of the autocorrelation is in essence then the read out when the PD voltage is displayed on a spectrum analyzer. For example on the Agilent N9030A, we read out the $V_{rms}(f)$ voltage and divide by the RMS noise bandwidth value to obtain the voltage fluctuation. For a Gaussian filter on the N9030A, the 3dB bandwidth is related to the noise bandwidth by $RBW_{noise} = 1.06 RBW_{3dB}$. RIN is commonly expressed in dB/Hz scale and can be calculated as

$$RIN \, [dB/Hz] = 10 \log_{10} \left( \frac{2 V_{rms}(f_m)^2 / RBW_{noise}}{\bar{V}^2} \right) \tag{3}$$

Note, other spectrum analyzers may be able to scale out the PSD automatically by choosing the appropriate y units. We set up a RIN measurement of the Brillouin and locked pump lasers as shown schematically in Fig. S10. We note that in order to not damage the Spectrum analyzer, we had to use a DC block which stopped frequencies below 10 kHz. The peaks at multiples of ~ 2.5 MHz correspond to the spacing of the Brillouin cavity and may be due to a small amount of Stokes light generated at other fiber cavity resonances. We teed off the PD signal to measure the average DC voltage which corresponds to 3.547 V. For comparison, we disconnected the fiber input to the photodiode and performed a similar calculation to measure the electronic noise floor of this measurement. We recorded the RIN spectra with the N9030A spectrum analyzer, all measurements were recorded with an RBW of 1 Hz. Data was stitched together using a 1000 points between each decade of frequency. All data was averaged over 15 traces except for the 1 − 10 MHz decade where only 5 averages were used due to time constraints (Fig. S10).

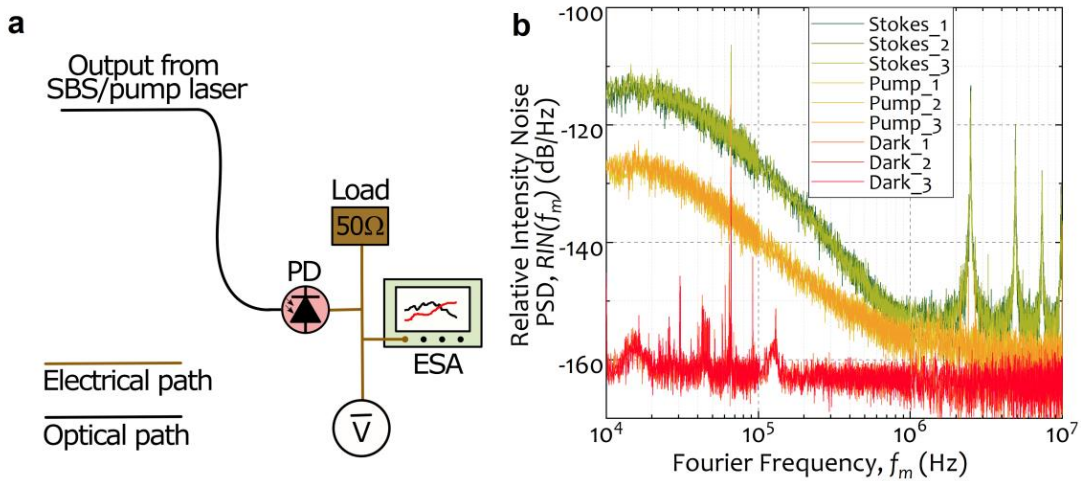

**Fig. S10: RIN measurement. (a)** Schematic of the RIN measurement apparatus. **(b)** Measured and calibrated RIN power spectral densities (PSD) for Fourier frequencies ranging from 10 kHz to 10 MHz for multiple measurements.

## 10. Supporting Video Discussion

The supporting video confirms continuous tuning of the pump –Stokes beat corresponding to pump laser wavelengths from 1549.054 nm to 1549.598 nm. This range reflects the pump drive current, $I_{LD}$ ranges from 48.6 mA to 78.8 mA. The electrical spectrum analyzer (ESA) on the top display shows the RF beat signal from a fast photodiode response while the bottom optical spectrum analyzer (OSA) shows the same beat in the optical domain, i.e. the combined pump and Stokes signal. The span of the OSA in the video is 1 nm and continuous tuning of the resultant combined pump – Stokes signal can be observed for more than 0.5 nm before the strong beat signal in the ESA crumbles down. Slight effects of noisier spectrum are observed in the ESA within the continuous wavelength tuning range which can be attributed to the free spectral range (FSR) of the Brillouin cavity between resonances in wavelength. This continuous wavelength tuning is facilitated by keeping the synthesizer frequency fixed at 10.876 GHz while utilizing a higher output power (8 dBm). Beyond the continuous tuning range at the stated frequency and power combination, the strong beat signal in the ESA breaks down to multiple minor beat signals corresponding to the frequency difference between the necessary changed Brillouin shift corresponding to the refreshed wavelength arising from the tuning and the operating synthesizer frequency.

## 11. Two Wavelength ($2-\lambda$) Difference Frequency Measurement

We assembled an electro–optic (EO) comb comprising of three electro-optic phase modulators connected in tandem along with their respective high current amplifiers (**Fig. S11a**). In this configuration, the phase modulators can generate sidebands in operation where cascading effect causes the first modulator central carrier and sidebands to result in extra modulator sidebands for the next phase modulator. These multitude of carrier and sidebands are then input for the third modulator which again replicates this effect and ends up in furthermore sidebands upon modulation, similar to a sort of avalanching effect. The general scheme of this type of comb consists of a continuous-wave (CW) laser sent to a set of electro-optic modulators driven by an external RF oscillator, which is a dielectric resonant oscillator (DRO) in this case. The modulation introduces sidebands around the optical frequency defined by the input laser. Hence, the performance of this platform is dictated by the noise characteristics of the laser, oscillator, optical components, and microwave amplifiers. This solution combines robustness and simplicity at a few nm bandwidth (considering ∼10 GHz oscillator and state-of-the-art optoelectronic and microwave components).

We channeled the two wavelength Stokes light, i.e. system output in Fig. 4 of the main text, into the EO comb for sideband generation and subsequently through an optical bandpass filter (OBPF) to filter out the central beat based on the number of sidebands and the two wavelength difference (TWD) frequency in question (Fig. S11a). The intermediate frequency of the central sideband beat note is then photo-detected. The only drawback in this process is that the DRO noise is also added to the base noise of the Stokes signal and also that only TWD frequencies near $q \cdot 2f_{DRO}$ can be measured, where $q$ is an integer. The DRO noise dominates the measured phase noise PSDs of the TWD high frequency signals ranging from 180 – 500 GHz (Fig. S11, c – f). This fact can be ascertained from the multiplied up DRO phase noise PSDs closely following the corresponding phase noise PSDs of all the TWD signals in the high frequency range 108 GHz – 500 GHz.

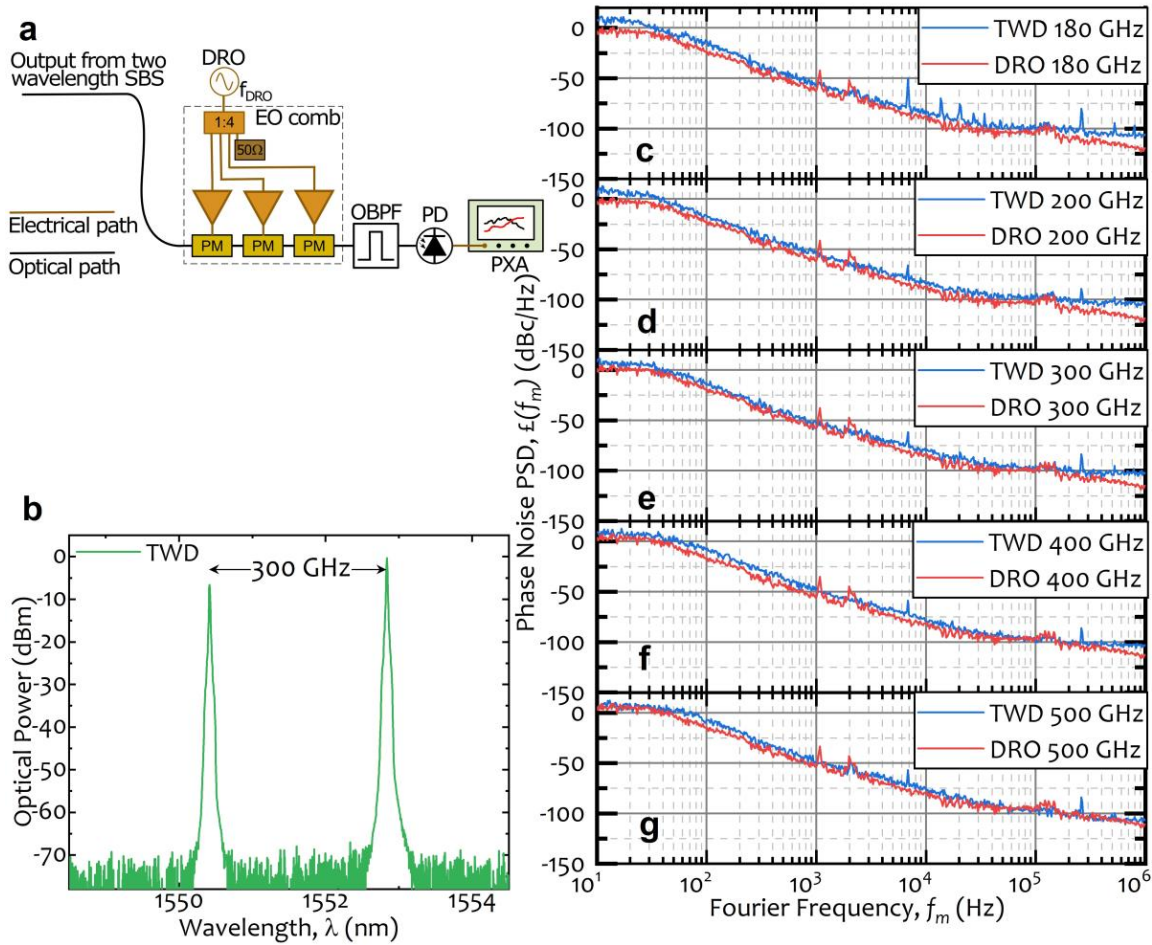

**Fig. S11: Two wavelength difference (TWD) frequency measurements. (a)** Measurement system for the two wavelength difference high frequency beats. **(b)** Optical spectrum of the TWD frequency corresponding to 300 GHz separation. **(c – g)** Measured two wavelength difference (TWD) high frequency beat note phase noises along with associated DRO noise multiplied up to that frequency range. The multiplied dielectric resonant oscillator (DRO) noise closely follows the measured phase noise power spectral densities (PSD) of the two wavelength difference high frequency beat notes.

## 12. Interferometric Characterization of the Generated mmW Signal

When the separation between the two SBS lasers are 500 GHz, the two Stokes lines exhibit about 7 dB difference in optical power out of the Brillouin cavity due to the different pumps being offset by unequal amounts from their respective center operating wavelengths. The opportunistic filtering with optical band OBPFs and selective routing of SBS laser lines through AOM1 in Fig. 3a(i) in the main text reduces this optical power imbalance to about 2 dB after the EDFA (Fig. S12a).

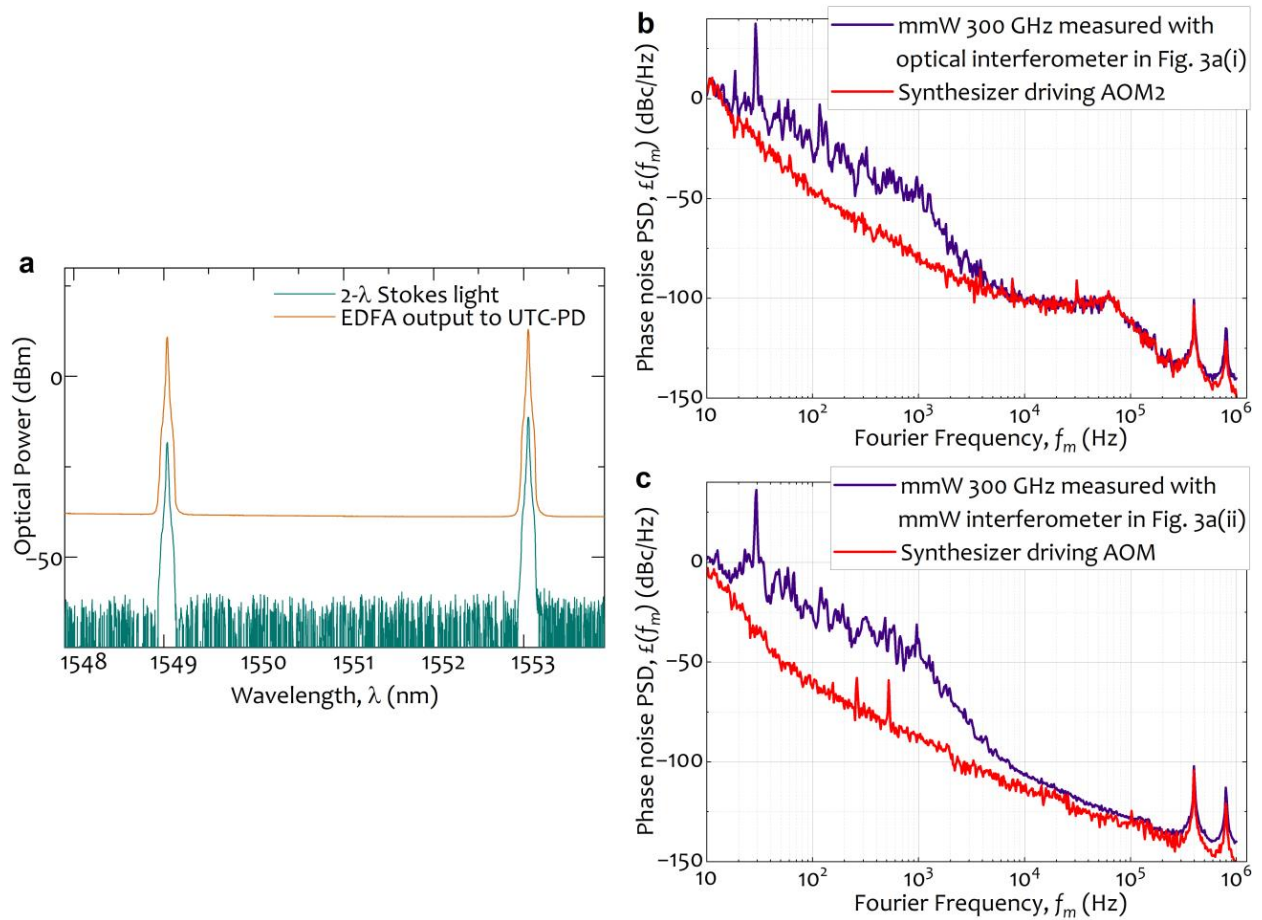

**Fig. S12: Millimeter wave (mmW) characterization details.** (a) Optical spectra of the $2-\lambda$ Stokes signal and EDFA output before the UTC-PD (in Fig. 3a(i) in the main text) corresponding to a wavelength difference of 0.5 THz. **(b, c)** The measured phase noise power spectral densities (PSD) of the two wavelength difference beat notes at 300 GHz for the two employed characterization schemes in Fig. 3a[(i) and (ii)] in the main text and corresponding synthesizer (source) phase noises limiting the respective measurements.

The interferometric approach to phase noise characterization of the 300 GHz wave is limited by the source synthesizer noise in both cases of measurements – Fig. 3a(i) and Fig. 3a(ii). The phase noise for the synthesizer driving the AOM2 (Rigol DG4102) in Fig. 3a(i) at 82 MHz is measured and calibrated for the delay line of 500 m and plotted with the phase noise achieved from the setup in Fig. 3a(i) for the 300 GHz wave (Fig. S12b). The synthesizer noise limits the achieved phase noise levels for Fourier frequencies at nearly 10 kHz and above. The mmW frequency mixing provided improved results in terms of phase noise, however, still limited by the synthesizer driving the AOM in Fig. 3a(ii) in the main text. We used a low noise synthesizer (Agilent N5181A) and divided the frequency with a frequency divider (Valon 3010a) to obtain a frequency supported by the AOM. The phase noise for the divided frequency is measured

(Fig. S12c). Still, the achieved phase noise levels with the mmW frequency mixing approach (Fig. 3a(ii) in the main text) is limited by the source phase noise for Fourier frequencies above 100 kHz. This suggests that the measurement noise floor at those frequencies are limited by our source and the actual mmW signal could even fetch lower white phase noise floors.